\documentclass[12pt]{article}
\usepackage{lscape}
\usepackage{citesort}
\usepackage{amssymb}
\usepackage{appendix}
\usepackage{multirow}

\begin{document}

\def\qq{\langle \bar q q \rangle}
\def\uu{\langle \bar u u \rangle}
\def\dd{\langle \bar d d \rangle}
\def\sp{\langle \bar s s \rangle}
\def\GG{\langle g_s^2 G^2 \rangle}
\def\Tr{\mbox{Tr}}
\def\figt#1#2#3{
        \begin{figure}
        $\left. \right.$
        \vspace*{-2cm}
        \begin{center}
        \includegraphics[width=10cm]{#1}
        \end{center}
        \vspace*{-0.2cm}
        \caption{#3}
        \label{#2}
        \end{figure}
	}
	
\def\figb#1#2#3{
        \begin{figure}
        $\left. \right.$
        \vspace*{-1cm}
        \begin{center}
        \includegraphics[width=10cm]{#1}
        \end{center}
        \vspace*{-0.2cm}
        \caption{#3}
        \label{#2}
        \end{figure}
                }

\def\ds{\displaystyle}
\def\beq{\begin{equation}}
\def\eeq{\end{equation}}
\def\bea{\begin{eqnarray}}
\def\eea{\end{eqnarray}}
\def\beeq{\begin{eqnarray}}
\def\eeeq{\end{eqnarray}}
\def\ve{\vert}
\def\vel{\left|}
\def\ver{\right|}
\def\nnb{\nonumber}
\def\ga{\left(}
\def\dr{\right)}
\def\aga{\left\{}
\def\adr{\right\}}
\def\lla{\left<}
\def\rra{\right>}
\def\rar{\rightarrow}
\def\lrar{\leftrightarrow}  
\def\nnb{\nonumber}
\def\la{\langle}
\def\ra{\rangle}
\def\ba{\begin{array}}
\def\ea{\end{array}}
\def\tr{\mbox{Tr}}
\def\ssp{{\Sigma^{*+}}}
\def\sso{{\Sigma^{*0}}}
\def\ssm{{\Sigma^{*-}}}
\def\xis0{{\Xi^{*0}}}
\def\xism{{\Xi^{*-}}}
\def\qs{\la \bar s s \ra}
\def\qu{\la \bar u u \ra}
\def\qd{\la \bar d d \ra}
\def\qq{\la \bar q q \ra}
\def\gGgG{\la g^2 G^2 \ra}
\def\GG{\langle g_s^2 G^2 \rangle}
\def\g5{\gamma_5 \not\!q}
\def\x{\gamma_5 \not\!x}
\def\g5{\gamma_5}
\def\sb{S_Q^{cf}}
\def\sd{S_d^{be}}
\def\su{S_u^{ad}}
\def\sbp{{S}_Q^{'cf}}
\def\sdp{{S}_d^{'be}}
\def\sup{{S}_u^{'ad}}
\def\ssp{{S}_s^{'??}}

\def\sig{\sigma_{\mu \nu} \gamma_5 p^\mu q^\nu}
\def\fo{f_0(\frac{s_0}{M^2})}
\def\ffi{f_1(\frac{s_0}{M^2})}
\def\fii{f_2(\frac{s_0}{M^2})}
\def\O{{\cal O}}
\def\sl{{\Sigma^0 \Lambda}}
\def\es{\!\!\! &=& \!\!\!}
\def\ap{\!\!\! &\approx& \!\!\!}
\def\ar{&+& \!\!\!}
\def\arrr{\!\!\!\! &+& \!\!\!}
\def\ek{&-& \!\!\!}
\def\vev{&\vert& \!\!\!}
\def\kek{\!\!\!\!&-& \!\!\!}
\def\cp{&\times& \!\!\!}
\def\se{\!\!\! &\simeq& \!\!\!}
\def\eqv{&\equiv& \!\!\!}
\def\kpm{&\pm& \!\!\!}
\def\kmp{&\mp& \!\!\!}
\def\mcdot{\!\cdot\!}
\def\erar{&\rightarrow&}


\def\simlt{\stackrel{<}{{}_\sim}}
\def\simgt{\stackrel{>}{{}_\sim}}


\renewcommand{\textfraction}{0.2}    
\renewcommand{\topfraction}{0.8}   

\renewcommand{\bottomfraction}{0.4}   
\renewcommand{\floatpagefraction}{0.8}
\newcommand\mysection{\setcounter{equation}{0}\section}

\def\baeq{\begin{appeq}}     \def\eaeq{\end{appeq}}  
\def\baeeq{\begin{appeeq}}   \def\eaeeq{\end{appeeq}}
\newenvironment{appeq}{\beq}{\eeq}   
\newenvironment{appeeq}{\beeq}{\eeeq}
\def\bAPP#1#2{
 \markright{APPENDIX #1}
 \addcontentsline{toc}{section}{Appendix #1: #2}
 \medskip
 \medskip
 \begin{center}      {\bf\LARGE Appendix #1 }{\quad\Large\bf #2}
\end{center}
 \renewcommand{\thesection}{#1.\arabic{section}}
\setcounter{equation}{0}
        \renewcommand{\thehran}{#1.\arabic{hran}}
\renewenvironment{appeq}
  {  \renewcommand{\theequation}{#1.\arabic{equation}}
     \beq
  }{\eeq}
\renewenvironment{appeeq}
  {  \renewcommand{\theequation}{#1.\arabic{equation}}
     \beeq
  }{\eeeq}
\nopagebreak \noindent}

\def\eAPP{\renewcommand{\thehran}{\thesection.\arabic{hran}}}

\renewcommand{\theequation}{\arabic{equation}}
\newcounter{hran}
\renewcommand{\thehran}{\thesection.\arabic{hran}}

\def\bmini{\setcounter{hran}{\value{equation}}
\refstepcounter{hran}\setcounter{equation}{0}
\renewcommand{\theequation}{\thehran\alph{equation}}\begin{eqnarray}}
\def\bminiG#1{\setcounter{hran}{\value{equation}}
\refstepcounter{hran}\setcounter{equation}{-1}
\renewcommand{\theequation}{\thehran\alph{equation}}
\refstepcounter{equation}\label{#1}\begin{eqnarray}}


\newskip\humongous \humongous=0pt plus 1000pt minus 1000pt
\def\caja{\mathsurround=0pt}
 

\title{
         {\Large
                 {\bf
Strong coupling constant of negative parity nucleon with $\pi$ meson 
in light cone QCD sum rules
                 }
         }
      }

\author{\vspace{1cm}\\
{\small T. M. Aliev$^1$ \thanks {e-mail: taliev@metu.edu.tr}~\footnote{permanent address:Institute of
Physics,Baku,Azerbaijan}\,\,,
T. Barakat$^2$ \thanks {e-mail:
tbarakat@KSU.EDU.SA}\,\,,
M. Savc{\i}$^1$ \thanks
{e-mail: savci@metu.edu.tr}} \\
{\small $^1$\,Physics Department, Middle East Technical University,
06800 Ankara, Turkey }\\
{\small $^2$\,Physics and Astronomy Department, King Saud University, Saudi
Arabia}}

\date{}

\begin{titlepage}
\maketitle
\thispagestyle{empty}

\begin{abstract}

We estimate strong coupling constant between the negative parity nucleons
with $\pi$ meson within the light cone QCD sum rules. A method for
eliminating the unwanted contributions coming from the nucleon--nucleon and
nucleon--negative parity nucleon transition is presented. It is
observed that the value strong coupling constant of the negative parity nucleon 
$N^\ast N^\ast \pi$ transition is considerably different from the one predicted by the
3--point QCD sum rules, but is quite close to the coupling constant of the
positive parity $N N \pi$ transition.

\end{abstract}

~~~PACS numbers: 14.20.Dh, 14.40.Be, 13.75.Gx, 11.55.Hx

\end{titlepage}

\section{Introduction}

The strong coupling constants of hadrons with mesons are the key quantity
for understanding the dynamics of the existing hadron--hadron,
hadron--meson, and photoproduction experiments. Among many couplings, only
nucleon--pion coupling constant has been measured accurately from
experiments. With increasing experimental information there appears
the necessity for an accurate determination of the strong coupling
constants of hadrons with pseudoscalar mesons. These coupling constants have
so far been estimated within various approaches (relevant references can be
found in \cite{Rfts01}).
The strong coupling constants
of the octet baryons with pseudoscalar mesons are calculated in framework of
the light cone QCD sum rules \cite{Rfts01}.

In the present note we calculate the $N^\ast N^\ast \pi$ coupling constant in light
cone QCD sum rules. Compared to the all other sum rules approaches that take
only one positive party baryon into consideration, the main novelty of the
present calculation is that the contributions to the sum rules coming from
the two positive parity $N(938)$ and $N(1440)$ states are taken into
account. This fact makes the analysis of the sum rules more complicated in
determination of the $N^\ast N^\ast \pi$ coupling constant.
In the present work a new method is explored for eliminating 
the unwanted contributions coming from the $N(938) \to N(938)$,
$N(1440) \to N(1440)$, $N^\ast \to N(938)$, $N^\ast \to N(1440)$
transitions. In the following discussion we shall customarily denote
$N(1440)$ as $N^\prime$. It should be noted here that the $N^\ast N^\ast
\pi$ coupling constant is determined in \cite{Rfts02} in framework of the
3--point QCD sum rules.

The paper is organized as follows. In section 2 we derive the sum rules for
the $N^\ast N^\ast \pi$ coupling constant. In section 3 we present the
numerical analysis of the sum rules for this parameter, and compare our
result  with the prediction of 3--point QCD sum rules.

\section{Sum rules for the $N^\ast N^\ast \pi$ coupling}

In this section sum rules for the $N^\ast N^\ast \pi$ coupling constant
within the light cone QCD sum rules method is derived. In determining this
coupling constant we consider the following correlation function,

\bea
\label{efts01}
\Pi(p,q) = i \int d^4x e^{ipx} \lla \pi(q) \vel J_N (x) J_N^\dag (0) \ver 0
\rra\,,
\eea
where
\bea
\label{efts02}
J_N = \varepsilon^{abc} \Big\{ \left(u^{a\dag} C d^b \right) \gamma_5 u^c +
\beta \left(u^{a\dag} C\gamma_5 d^b\right) u^c \Big\}\,,
\eea

is the interpolating current of the nucleon, $a,\,b,\,c$ are the color
indices, $C$ is the charge conjugation operator, and $\beta$ is an arbitrary
parameter. The sum rules for the $N^\ast N^\ast \pi$ coupling can be
obtained by following the QCD sum rules procedure. On the one hand, the
correlation function is calculated in terms of hadrons. On the other hand,
it can be calculated in the deep Eucledian domain
$p^2\!\ll \!0,~(p+q)^2 \!\ll \!0$
by using the operator product expansion (OPE) over twist. By matching then
these two representations, the sum rules for the $N^\ast N^\ast \pi$
coupling is obtained.

The hadronic part of the correlation function is obtained by inserting a
complete set of baryons, and isolating the ground state contributions of the
baryons as follows,
\bea
\label{efts03}
\Pi = \sum_{i,j} {\lla 0 \vel J_N \ver N_i(p,s) \rra \lla \pi(q) N_i \ve N_j
(p+q) \rra \lla N_j \vel \bar{J}_N \ver 0 \rra \over
(p_i^2-m_i^2)[(p+q)^2 - m_j^2] }\,,
\eea
where $i$ and $j$ run over $N$, $N(1440)$ and $N^\ast(1535)$, and dots
denote higher state contributions. The matrix elements in Eq. (\ref{efts03})
are defined as,
\bea
\label{efts04}  
\lla 0 \vel J_N\ver N(N^\prime) \rra \es \lambda_{N(N^\prime)}
u_{N(N^\prime)}(p)\,, \nnb \\
\lla \pi(q) N_i \ve N_j(p+q) \rra \es g_{N_i N_j\pi} \bar{u}_{N_i} \Gamma_j
u_{N_j}\,, \nnb\\
\lla 0 \vel J_N \ver N^\ast \rra \es \lambda_{N^\ast} \gamma_5 u_{N^\ast}
(p)\,,
\eea
where
\bea
\Gamma_j = \left\{ \begin{array}{l}
i\gamma_5,~\mbox{for }N\to N,~N^\prime \to N^\prime,~N^\prime \to N,~N^\ast
\to N^\ast~\mbox{transitions},\\
~~I,~\mbox{for }N^\ast \to N,~N^\ast \to N^\prime ~\mbox{transitions}.\\ 
\end{array} \right. \nnb 
\eea
Using Eqs. (\ref{efts03}) and (\ref{efts04}), for the hadronic part of the
correlation function we get,
\bea
\label{efts05}
\Pi \es
A (\not\!{p} + m_N) i\gamma_5 (\not\!{p} + \not\!{q} + m_N) +
B (\not\!{p} + m_{N^\prime}) i\gamma_5 (\not\!{p} + \not\!{q} + m_{N^\prime}) \nnb \\
\ar C (\not\!{p} + m_{N^\prime}) i\gamma_5 (\not\!{p} + \not\!{q} + m_N) +
D (\not\!{p} + m_N) i\gamma_5 (\not\!{p} + \not\!{q} + m_{N^\prime}) \nnb \\  
\ar E \gamma_5 (\not\!{p} + m_{N^\ast}) i\gamma_5 (\not\!{p} + \not\!{q} +
m_{N^\ast}) (-\gamma_5) +
F i\gamma_5 (\not\!{p} + m_{N^\ast}) (\not\!{p} + \not\!{q} + m_N) \nnb \\
\ar K i \gamma_5 (\not\!{p} + m_{N^\ast}) (\not\!{p} + \not\!{q} +
m_{N^\prime}) +
L (\not\!{p} + m_N) (\not\!{p} + \not\!{q} + m_{N^\ast}) (-i\gamma_5)  \nnb \\        
\ar N (\not\!{p} + m_{N^\prime}) (\not\!{p} + \not\!{q} + m_{N^\ast})
(-i\gamma_5)\,,
\eea
where
\bea
\label{efts06}
A \es {\vel \lambda_N \ver^2 g_{NN\pi}\over (p^2-m_N^2) [(p+q)^2 - m_N^2]}\,,\nnb \\
B \es {\vel \lambda_{N^\prime}\ver^2  g_{N^\prime N^\prime \pi} \over (p^2-m_{N^\prime}^2)[(p+q)^2 -
m_{N^\prime}^2]}\,,\nnb \\
C \es { \lambda_N^\ast \lambda_{N^\prime} g_{N N^\prime \pi}\over (p^2-m_{N^\prime}^2)[(p+q)^2 -  
m_N^2]}\,,\nnb \\
D \es { \lambda_{N^\prime}^\ast \lambda_N g_{N^\prime N \pi}\over (p^2-m_N^2)[(p+q)^2 -        
m_{N^\prime}^2)]}\,,\nnb \\
E \es {\vel \lambda_{N^\ast} \ver^2 g_{N^\ast N^\ast \pi} \over (p^2-m_{N^\ast}^2) [(p+q)^2 -
m_{N^\ast}^2]}\,,\nnb \\
F \es { \lambda_{N^\ast}^\ast \lambda_N g_{N^\ast N \pi} \over (p^2-m_{N^\ast}^2) [(p+q)^2 -
m_N^2]}\,,\nnb \\  
K \es { \lambda_{N^\ast}^\ast \lambda_{N^\prime} g_{N^\ast N^\prime \pi}\over (p^2-m_{N^\ast}^2) [(p+q)^2 -     
m_{N^\prime}^2]}\,,\nnb \\
L \es { \lambda_N^\ast \lambda_{N^\ast} g_{N N^\ast \pi}\over (p^2-m_N^2)
[(p+q)^2 - m_{N^\ast}^2]}\,,\nnb \\
N \es { \lambda_{N^\prime}^\ast \lambda_{N^\ast} g_{N^\prime N^\ast \pi}\over (p^2-m_{N^\prime}^2)
[(p+q)^2 - m_{N^\ast}^2]}\,.
\eea      

The correlation function can be calculated from the QCD side by using Wick's
theorem. In calculation of the correlation function from theoretical side
the expression of the light quark operators in the presence of an external
field are needed, and it is calculated in \cite{Rfts03}. It should be noted
here that the quark propagator gets contributions from three--particle
$\bar{q}Gq$, four--particle $\bar{q}G^2 q$, $\bar{q}q\bar{q}q$ nonlocal
operators. In further calculations we take into account only three--particle
$\bar{q}Gq$ operator and neglect contributions coming from four--particle
operators. It is demonstrated in  \cite{Rfts03} that neglecting these
contributions can be legitimated on the basis of an expansion over conformal
spin. Under this approximation the light quark propagator in the background
field is given by,
\bea
\label{efts07}
S_q(x) \es {i \rlap/x\over 2\pi^2 x^4} - {m_q\over 4 \pi^2 x^2}
- {i g_s \over 16 \pi^2 x^2} \int_0^1 du \Bigg[\rlap/{x} \bar{u}
\sigma_{\alpha \beta} G^{\alpha \beta} (ux)
+ u \sigma_{\alpha \beta} G^{\alpha \beta} (ux) \nnb \\
\ek {1\over 2} i m_q x^2 \sigma_{\alpha \beta} G^{\alpha \beta} (ux)
\left( \ln {-x^2 \Lambda^2\over 4}  +
 2 \gamma_E \right) \Bigg]\,,
\eea
where $\Lambda$ is the parameter separating the perturbative and
nonperturbative domains, whose value is estimated to be $\Lambda=(0.5 \div
1.0)\,GeV$ in \cite{Rfts04}.

Using the expression of the light quark propagator, the correlation function
can be calculated from the QCD side straightforwardly in deep Eucledian
region $p^2 \to - \infty$, $(p+q)^2 \to - \infty$ by using the operator
product expansion over twist. In this calculation the matrix elements of the
nonlocal operators $\bar{q}(x) \Gamma q(0)$ and $\bar{q}(x) \Gamma
G_{\mu\nu} (ux) q(0)$ between the vacuum and pion states appear, where
$\Gamma$ corresponds to the matrices from full set of Dirac matrices.
The matrix elements up to twist--4 are parametrized in terms of the pion
distribution amplitudes in the following way
\cite{Rfts05,Rfts06,Rfts07,Rfts08,Rfts09}:
\bea
\label{efts08} 
\lla \pi(p)\vel \bar q_1(x) \gamma_\mu \gamma_5 q_1(0)\ver 0 \rra \es
-i f_\pi q_\mu  \int_0^1 du  e^{i \bar u q x}
    \left( \varphi_\pi(u) + {1\over 16} m_\pi^2
x^2 {\Bbb{A}}(u) \right) \nnb \\
\ek {i\over 2} f_\pi m_\pi^2 {x_\mu\over qx}
\int_0^1 du e^{i \bar u qx} {\Bbb{B}}(u)\,,\nnb \\
\lla \pi(p)\vel \bar q_1(x) i \gamma_5 q_2(0)\ver 0 \rra \es
\mu_\pi \int_0^1 du e^{i \bar u qx} \phi_P(u)\,,\nnb \\
\lla \pi(p)\vel \bar q_1(x) \sigma_{\alpha \beta} \gamma_5 q_2(0)\ver 0 \rra \es
{i\over 6} \mu_\pi \left( 1 - \widetilde{\mu}_\pi^2 \right)
\left( q_\alpha x_\beta - q_\beta x_\alpha\right)
\int_0^1 du e^{i \bar u qx} \phi_\sigma(u)\,,\nnb \\
\lla \pi(p)\vel \bar q_1(x) \sigma_{\mu \nu} \gamma_5 g_s
G_{\alpha \beta}(v x) q_2(0)\ver 0 \rra \es i \mu_\pi \left[
q_\alpha q_\mu \left( g_{\nu \beta} - {1\over qx}(q_\nu x_\beta +
q_\beta x_\nu) \right) \right. \nnb \\
\ek q_\alpha q_\nu \left( g_{\mu \beta} -
{1\over qx}(q_\mu x_\beta + q_\beta x_\mu) \right) \nnb \\
\ek q_\beta q_\mu \left( g_{\nu \alpha} - {1\over qx}
(q_\nu x_\alpha + q_\alpha x_\nu) \right) \nnb \\
\ar q_\beta q_\nu \left. \left( g_{\mu \alpha} -
{1\over qx}(q_\mu x_\alpha + q_\alpha x_\mu) \right) \right] \nnb \\
\cp \int {\cal D} \alpha e^{i (\alpha_{\bar q} +
v \alpha_g) qx} {\cal T}(\alpha_i)\,,\nnb \\
\lla \pi(p)\vel \bar q_1(x) \gamma_\mu \gamma_5 g_s
G_{\alpha \beta} (v x) q_2(0)\ver 0 \rra \es q_\mu (q_\alpha x_\beta -
q_\beta x_\alpha) {1\over qx} f_\pi m_\pi^2
\int {\cal D}\alpha e^{i (\alpha_{\bar q} + v \alpha_g) qx}
{\cal A}_\parallel (\alpha_i) \nnb \\
\ar \left[q_\beta \left( g_{\mu \alpha} - {1\over qx}
(q_\mu x_\alpha + q_\alpha x_\mu) \right) \right. \nnb \\
\ek q_\alpha \left. \left(g_{\mu \beta}  - {1\over qx}
(q_\mu x_\beta + q_\beta x_\mu) \right) \right]
f_\pi m_\pi^2 \nnb \\
\cp \int {\cal D}\alpha e^{i (\alpha_{\bar q} + v \alpha _g)
q x} {\cal A}_\perp(\alpha_i)\,,\nnb \\
\lla \pi(p)\vel \bar q_1(x) \gamma_\mu i g_s G_{\alpha \beta}
(v x) q_2(0)\ver 0 \rra \es q_\mu (q_\alpha x_\beta - q_\beta x_\alpha)
{1\over qx} f_\pi m_\pi^2 \int {\cal D}\alpha e^{i (\alpha_{\bar q} +
v \alpha_g) qx} {\cal V}_\parallel (\alpha_i) \nnb \\
\ar \left[q_\beta \left( g_{\mu \alpha} - {1\over qx}
(q_\mu x_\alpha + q_\alpha x_\mu) \right) \right. \nnb \\
\ek q_\alpha \left. \left(g_{\mu \beta}  - {1\over qx}
(q_\mu x_\beta + q_\beta x_\mu) \right) \right] f_\pi m_\pi^2 \nnb \\
\cp \int {\cal D}\alpha e^{i (\alpha_{\bar q} +
v \alpha _g) q x} {\cal V}_\perp(\alpha_i)\,,
\eea
where
\bea
\mu_\pi = f_\pi {m_\pi^2\over m_u + m_d}\,,~~~~~
\widetilde{\mu}_\pi = {m_u + m_d \over m_\pi}\,, \nnb
\eea
and ${\cal D}\alpha = d\alpha_{\bar q} d\alpha_q d\alpha_g
\delta(1-\alpha_{\bar q} - \alpha_q - \alpha_g)$.  
In these expressions $\varphi_\pi(u)$ is the leading twist--two, $\phi_\pi(u)$,
$\phi_\sigma(u)$, ${\cal T}(\alpha_i)$ are the twist--three, and
remaining ones are twist--four DAs, respectively, whose explicit expressions
are given in the following section.

It follows from Eq. (\ref{efts05}) that we have four independent Lorentz structures 
$\rlap/{p}\rlap/{q}\gamma_5$, $\rlap/{q}\gamma_5$, $ \rlap/{p}\gamma_5$ and
$\gamma_5$ for the problem under consideration. In principle,
the relevant sum rules can be obtained by performing double Borel
transformation over the variables $-p^2$ and $-(p+q)^2$ on the theoretical
and hadronic parts, and matching the coefficients of the corresponding
Lorentz structures. At this point we face the following problem.
Among the nine coefficients given in Eq. (\ref{efts07})
only coefficient $E$ describes the $N^\ast N^\ast \pi$ coupling constant.
However, as has already been noted, we have only four independent Lorentz
structures, and we need additional five more equations to determine the
$N^\ast N^\ast \pi$ coupling constant uniquely. Four of these additional five
equations can be obtained by taking derivatives of the four equations with
respect to the inverse Borel mass square. The fifth equation is obtained by
taking second derivative of the coefficient of the structure
$\rlap/{p}\rlap/{q}\gamma_5$ with respect to again the inverse Borel mass square.
From the numerical solution of these nine equations we get,
\bea
\label{efts09}
g_{N^\ast N^\ast \pi} \vel \lambda_{N^\ast} \ver^2 e^{-m_{N^\ast}^2/M^2} \es
0.097 \Pi_1^B + 0.029 \Pi_2^B - 0.178
\Pi_3^B - 0.143 \Pi_4^B - 0.116 \Pi_1^{B\prime} \nnb \\
\ek 0.051 \Pi_2^{B\prime} + 0.062 \Pi_3^{B\prime} - 0.013 \Pi_4^{B\prime}
+ 0.024 \Pi_1^{B\prime \prime} +\cdots
\eea
where $\Pi_1^B$, $\Pi_2^B$, $\Pi_3^B$, and $\Pi_4^B$ are the coefficients of the
structures $\rlap/{p}\rlap/{q}\gamma_5$, $\rlap/{q}\gamma_5$, $
\rlap/{p}\gamma_5$ and $\gamma_5$ after Borel transformations with respect
to the variables $-(p+q)^2$ and $-p^2$ are performed, respectively;
$\Pi_i^{B\prime}$ stands for
the first derivative of $\Pi_i^B$ with respect to $1/M_1^2$, i.e.,
$d\Pi_i^B/d(1/M_1^2)$, and $\Pi_1^{B\prime \prime}$ is the second derivative of
$\Pi_1^B$ with respect to $1/M_1^2$. Here we set $m_\pi^2=0$,
and dots correspond to contributions of continuum and
higher states. These contributions can be calculated by using the hadron--quark
duality, i.e., above some threshold in the $(s_1,s_2)$ plane the hadronic
spectral density is equal to the quark spectral density.  
Note that after taking derivatives of the invariant functions
we set $M_1^2=M_2^2=2 M^2$. The expressions of the functions $\Pi_1^B$, $\Pi_2^B$,
$\Pi_3^B$ and $\Pi_4^B$ are quite lengthy and for this reason we do not
present them in the present work.
Once Fourier and Borel
transformations are carried out, continuum subtraction can be performed by
using the following formula,
\bea
M^{2n} \to {1\over \Gamma(n)} \int_0^{s_0} ds\,e^{-s/M^2}
(s-n_{N^\ast}^2)^{n-1}\,,\nnb
\eea
which leads to
\bea
M^2 \to M^2\left( 1- e^{-s_0/M^2} \right)\,.\nnb
\eea

For the higher twist terms that are proportional to the negative power of
$M^2$, the subtraction procedure is not performed, since their contributions
are negligibly small (for more details, see \cite{Rfts05}).

Our final remark in this section is about the residue of the negative parity
baryons, which is determined from the two--point correlation function,
\bea
\label{efts10}
\Pi(p^2) = i \int d^4x e^{ipx} \lla 0 \vel \mbox{T}\Big( \bar{\eta}_N(x)
\eta_N (0) \Big) \ver 0 \rra \,,
\eea
where $\eta$ is given in Eq. (\ref{efts02}). Saturating this correlation
function with positive and negative parity baryons we get,
\bea
\label{efts11}
\Pi(p^2) \es { \vel \lambda_N \ver^2 (\not\!{p}+m_N) \over (p^2-m_N^2)}
+ { \vel \lambda_{N^\prime} \ver^2 (\not\!{p}+m_{N^\prime}) \over
(p^2-m_{N^\prime}^2)} +
{ \vel \lambda_{N^\ast} \ver^2 (\not\!{p}+m_{N^\ast}) \over
(p^2-m_{N^\ast}^2)}\,.
\eea

When we calculate this correlation function from theoretical side we get,
\bea
\label{efts12}
\Pi(p^2) \es \not\!{p} \Pi_1(p^2) I \Pi_2(p^2)\,,
\eea
where $\Pi_i$ are the corresponding invariant functions. Performing Borel
transformation over $p^2$, and equating the coefficients of the structures we
get,
\bea
\label{efts13}
\vel \lambda_N \ver^2 e^{-m_N^2/M^2} + \vel \lambda_{N^\prime} \ver^2
e^{-m_{N^\prime}^2/M^2} + \vel \lambda_{N^\ast} \ver^2
e^{-m_{N^\ast}^2/M^2} \es \widetilde{\Pi}_1^B\,,\nnb\\
m_N \vel \lambda_N \ver^2 e^{-m_N^2/M^2} + m_{N^\prime} \vel
\lambda_{N^\prime} \ver^2 e^{-m_{N^\prime}^2/M^2}
- m_{N^\ast} \vel \lambda_{N^\ast} \ver^2 e^{-m_{N^\ast}^2/M^2} \es
\widetilde{\Pi}_2^B\,.
\eea
Expressions of the invariant functions $\widetilde{\Pi}_1^B$ and
$\widetilde{\Pi}_2^B$ are given in \cite{Rfts10}.

As can easily be seen from these equations there are six unknowns, namely,
$m_N$, $\lambda_N$, $m_{N^\prime}$, $\lambda_{N^\prime}$,
$m_N^\ast$ and $\lambda_{N^\ast}$, and therefore we need six equations to be
able to solve for these unknown parameters. Two of these equations are given
in Eq. (\ref{efts13}), and the remaining four equations can be obtained from
Eq. (\ref{efts13}) by taking first and second derivatives with respect to
$(-1/M^2)$. Solving then these six equations we can determine
$\lambda_{N^\ast}$. Our numerical analysis shows that $\vel \lambda_{N^\ast}
\ver^2$ is positive in the regions $-1.0 \le
\cos\theta \le -0.8$ (where $\beta=\tan\theta$), and $0.8\le\cos\theta
\le 1.0$, and it is unphysical for all other values of $\cos\theta$.
therefore in further numerical analysis we will use these domains in
determination of the $N^\ast N^\ast \pi$ coupling constant.

\section{Numerical analysis}

In this section numerical analysis for the sum rules of the
strong coupling constant $g_{N^\ast N^\ast \pi}$ obtained in the previous
section is performed. In this analysis the values of the input parameters,
as well as the expressions of the pion distribution amplitudes (DAs) are needed,
which are the main ingredients of the light cone QCD sum rules. The
expressions of the pion DAs are given as,
\cite{Rfts05,Rfts06,Rfts07,Rfts08,Rfts09}

\bea
\label{efts17}
\varphi_{\cal P}(u) \es 6 u \bar u \left[ 1 + a_1^{\cal P} C_1(2 u -1) +
a_2^{\cal P} C_2^{3/2}(2 u - 1) \right]\,,  \nnb \\
{\cal T}(\alpha_i) \es 360 \eta_3 \alpha_{\bar q} \alpha_q
\alpha_g^2 \left[ 1 + w_3 {1\over 2} (7 \alpha_g-3) \right]\,, \nnb \\
\phi_P(u) \es 1 + \left[ 30 \eta_3 - {5\over 2}
{1\over \mu_{\cal P}^2}\right] C_2^{1/2}(2 u - 1)\,,  \nnb \\
\ar \left( -3 \eta_3 w_3  - {27\over 20} {1\over \mu_{\cal P}^2} -
{81\over 10} {1\over \mu_{\cal P}^2} a_2^{\cal P} \right)
C_4^{1/2}(2u-1)\,, \nnb \\
\phi_\sigma(u) \es 6 u \bar u \left[ 1 + \left(5 \eta_3 - {1\over 2} \eta_3 w_3 -
{7\over 20}  \mu_{\cal P}^2 - {3\over 5} \mu_{\cal P}^2 a_2^{\cal P} \right)
C_2^{3/2}(2u-1) \right] \,, \nnb \\
{\cal V}_\parallel(\alpha_i) \es 120 \alpha_q \alpha_{\bar q} \alpha_g
\left( v_{00} + v_{10} (3 \alpha_g -1) \right) \,, \nnb \\
{\cal A}_\parallel(\alpha_i) \es 120 \alpha_q \alpha_{\bar q} \alpha_g
\left( 0 + a_{10} (\alpha_q - \alpha_{\bar q}) \right) \,, \nnb \\
{\cal V}_\perp (\alpha_i) \es - 30 \alpha_g^2\left[ h_{00}(1-\alpha_g) +
h_{01} (\alpha_g(1-\alpha_g)- 6 \alpha_q \alpha_{\bar q}) +
h_{10}(\alpha_g(1-\alpha_g) - {3\over 2} (\alpha_{\bar q}^2+
\alpha_q^2)) \right] \,, \nnb \\
{\cal A}_\perp (\alpha_i) \es 30 \alpha_g^2(\alpha_{\bar q} - \alpha_q)
\left[ h_{00} + h_{01} \alpha_g + {1\over 2} h_{10}(5 \alpha_g-3) \right] \,, \nnb \\
B(u)\es g_{\cal P}(u) - \varphi_{\cal P}(u) \,, \nnb \\
g_{\cal P}(u) \es g_0 C_0^{1/2}(2 u - 1) + g_2 C_2^{1/2}(2 u - 1) +
g_4 C_4^{1/2}(2 u - 1) \,, \nnb \\
\Bbb{A}(u) \es 6 u \bar u \left[{16\over 15} + {24\over 35} a_2^{\cal P}+
20 \eta_3 + {20\over 9} \eta_4 +
\left( - {1\over 15}+ {1\over 16}- {7\over 27}\eta_3 w_3 -
{10\over 27} \eta_4 \right) C_2^{3/2}(2 u - 1)  \right. \nnb \\
    \ar \left. \left( - {11\over 210}a_2^{\cal P} - {4\over 135}
\eta_3w_3 \right)C_4^{3/2}(2 u - 1)\right] \,, \nnb \\
\ar \left( -{18\over 5} a_2^{\cal P} + 21 \eta_4 w_4 \right)
\left[ 2 u^3 (10 - 15 u + 6 u^2) \ln u  \right. \nnb \\
\ar \left. 2 \bar u^3 (10 - 15 \bar u + 6 \bar u ^2) \ln\bar u +
u \bar u (2 + 13 u \bar u) \right]\,,
\eea
where $C_n^k(x)$ are the Gegenbauer polynomials, and
\bea
\label{efts18}
h_{00}\es v_{00} = - {1\over 3}\eta_4 \,, \nnb \\
a_{10} \es {21\over 8} \eta_4 w_4 - {9\over 20} a_2^{\cal P} \,, \nnb \\
v_{10} \es {21\over 8} \eta_4 w_4 \,, \nnb \\
h_{01} \es {7\over 4}  \eta_4 w_4  - {3\over 20} a_2^{\cal P} \,, \nnb \\
h_{10} \es {7\over 4} \eta_4 w_4 + {3\over 20} a_2^{\cal P} \,, \nnb \\
g_0 \es 1 \,, \nnb \\
g_2 \es 1 + {18\over 7} a_2^{\cal P} + 60 \eta_3  + {20\over 3} \eta_4 \,, \nnb \\
g_4 \es  - {9\over 28} a_2^{\cal P} - 6 \eta_3 w_3\,.
\eea
The values of the parameters $a_1^{\cal P}$, $a_2^{\cal P}$,
$\eta_3$, $\eta_4$, $w_3$, and $w_4$ entering Eq. (\ref{efts18}) are listed in
Table (\ref{param}) for the pseudoscalar $\pi$, $K$ and $\eta$ mesons.

\begin{table}[h]
\def\bos{\lower 0.25cm\hbox{{\vrule width 0pt height 0.7cm}}}
\begin{center}
\begin{tabular}{|c|c|c|}
\hline\hline
        & \bos  $\pi$   &   $K$ \\
\hline
$a_1^{\cal P}$  & \bos   0 &   0.050 \\
\hline
$a_2^{\cal P}~\mbox{(set-1)}$  & \bos   0.11  &   0.15 \\
\hline
$a_2^{\cal P}~\mbox{(set-2)}$  &  \bos  0.25  &   0.27 \\
\hline
$\eta_3$    & \bos  0.015 &   0.015 \\
\hline
$\eta_4$    & \bos  10    &   0.6 \\
\hline
$w_3$       & \bos  $-3$    &   $-3$ \\
\hline
$w_4$       & \bos  0.2   &   0.2 \\
\hline \hline
\end{tabular}
\end{center}
\caption{Parameters of the wave function calculated at the renormalization scale $\mu = 1 \,GeV$}
\label{param}
\end{table}

The sum rules for the $N^\ast N^\ast \pi$ coupling constant contains three
additional auxiliary parameters, namely Borel mass $M^2$, continuum
threshold $s_0$ and the arbitrary number $\beta$. Obviously, the result for
the $N^\ast N^\ast \pi$ coupling constant should be independent of these
parameters. This leads to the necessity to find such regions of these
parameters where the strong coupling constant does not depend on them.
This issue can be handled by the following procedure. The first attempt is
to find a such a region of $M^2$ at several predetermined fixed values
of $s_0$ and $\beta$ so that $N^\ast N^\ast \pi$ coupling constant is
independent of its variation. The lower bound of $M^2$ is determined from
the condition that higher twist contributions are less than the leading
twist contributions. The upper bound is obtained by requiring that higher
states and continuum contributions constitute, say, 40\% of the perturbative
contribution. These conditions are both satisfied if the Borel mass parameter
varies in the region $1.5\,GeV^2 \le M^2 \le 2.5\,GeV^2$. Note that this
working region of $M^2$ is also obtained from analysis of the magnetic
moment of negative parity baryons \cite{Rfts10}.

In Figs. (1) and (2) we present the dependence of the strong coupling constant
$g_{N^\ast N^\ast \pi}$ on the Borel parameter $M^2$ at the fixed values
of the auxiliary parameter $\beta = -0.5$, $-0.3$, $0.0$, $0.3$, $0.5$
and at two fixed values
of the continuum threshold $s_0=4.0\,GeV^2$ and $s_0=4.5\,GeV^2$, respectively.
It follows from these figures that $g_{N^\ast N^\ast \pi}$ shows rather stable
behavior to the variation of $M^2$ in its working region.

The continuum threshold is the other arbitrary arbitrary parameter of the
sum rules. This parameter is related to the energy of the first excited
state. Analysis of various sum rules shows that $\sqrt{s_0} = m_{ground} +
\Delta$, where $m_{ground}$ is the ground state mass, and $\Delta$ is the          
energy difference between ground and first excited states which varies in
the domain $0.3\,GeV \le \sqrt{s_0} \le 0.8\,GeV$. In the present analysis we
use the average value $\sqrt{s_0} = (m_{ground} + 0.5)\,GeV$.

We also studied the dependence of the $N^\ast N^\ast \pi$ coupling
constant on $s_0$, at four different values of the auxiliary parameter
$\beta = -0.5;-0.3;0.0;0.3,0.5$, and at two fixed values of the Borel mass
parameter $M^2 = 2.0\,GeV^2$ and $M^2 = 2.5\,GeV^2$. We observe that
$g_{N^\ast N^\ast \pi}$ is practically insensitive to the variations in
$s_0$. The total result changes about 5--6\%

The final stage of sum rules is to find such a region of $\beta$ where
$g_{N^\ast N^\ast \pi}$ be independent of the variation in $\beta$. The
arbitrary parameter varies in the domain $-\infty \le \beta \le +\infty$.
This infinitely large region can be mapped into a more restricted domain by
introducing the definition $\beta = \tan\theta$, by running $\theta$ in the
region $0 \le \cos\theta \le \pi$.

In Figs. (3) and (4) we present the dependence of the coupling constant
$g_{N^\ast N^\ast \pi}$ on $\cos\theta$, at two fixed values of the continuum
threshold $s_0=4.0\,GeV^2$ and $s_0=4.5\,GeV^2$, and at the fixed values
of the Borel mass parameter $M^2=(1.5,\,2.0,\,2.5)\,GeV^2$, respectively.
We find that the coupling constant $g_{N^\ast N^\ast \pi}$ is
weakly dependent to the variation of $\cos\theta$ in the region $-1.0 \le \cos\theta 
\le -0.85$. We also perform similar analysis at two more fixed values of
the continuum threshold, $s_0=4.2\,GeV^2$ and $s_0=4.8\,GeV^2$, which shows
that the result for $g_{N^\ast N^\ast \pi}$ changes at most 7--8\%.

Taking into account the uncertainties coming from input parameters entering
into the pion DAs, as well as from quark condensates, residues of $N^\ast$
and from the parameters $M^2$ and $s_0$, we finally get the following
result,
\bea
g_{N^\ast N^\ast \pi} = (10\pm 2)\,. \nnb
\eea
Note that our prediction on $N^\ast N^\ast \pi$ coupling is about 50\% larger
compared to that obtained in 3--point QCD sum rules \cite{Rfts02}. This can
be explained by the fact that in the limit $q\to 0$ the result predicted by
3--point QCD sum rules is not reliable (for more details, see \cite{Rfts16}.

Finally we compare our result on the $N^\ast N^\ast \pi$ strong coupling constant
with the predictions of the $N N \pi$ coupling constant for the positive
parity baryons. The $g_{N N \pi}$ coupling constant is calculated in various works 
and the results obtained are summarized in the table below,
\bea
g_{N N \pi} = \left\{ \begin{array}{ll}
12\pm 5&\mbox{\cite{Rfts11,Rfts12}}\,, \\
9.76\pm 2.04&\mbox{\cite{Rfts13}}\,, \\
13.3 \pm 1.2&\mbox{\cite{Rfts14}}\, \\
14 \pm 4&\mbox{\cite{Rfts01}}\,, \\
13.5 \pm 0.5&\mbox{\cite{Rfts15}}\,.\nnb
\end{array} \right. \nnb
\eea

When we compare our results on the strong coupling constants of negative
parity baryons with pion with those similar coupling constants for the
positive parity baryons, we observe that our predictions are quite close the
results exiting in literature for the positive parity nucleon pion coupling
constant. Small difference in the results can be attributed to the different
values of the input parameters, value of the residue, and continuum
threshold $s_0$.

In summary, we calculate the strong coupling constant of negative parity
baryons with pion in framework of the light cone QCD sum rules. The unwanted
contributions coming from positive--to--positive, and positive--to--negative
parity transformations are eliminated by constructing combination of sum
rules corresponding to different Lorentz structures. In the case of nucleons
the situation becomes more challenging due to the second positive parity
baryon $N^\prime (1440)$ in addition to the ground state $N(938)$. Our
prediction on $N^\ast N^\ast \pi$ coupling constant is in good agreement
with those results for the positive parity baryons existing in literature,
but considerably different from the value predicted by the 3--point QCD sum
rules method.

\newpage

\section*{Figure captions}
{\bf Fig. (1)} The dependence of the strong coupling constant $g_{N^\ast
N^\ast \pi}$ on the Borel parameter $M^2$, at the fixed value of the
continuum threshold $s_0=4.0\,GeV^2$, and several fixed values of 
the auxliary parameter $\beta$.\\\\
{\bf Fig. (2)} The same as Fig. (1), but at the fixed value of the
continuum threshold $s_0=4.5\,GeV^2$.\\\\
{\bf Fig. (3)} The dependence of the strong coupling constant $g_{N^\ast
N^\ast \pi}$ on $\cos\theta$, at the fixed value of the
continuum threshold $s_0=4.0\,GeV^2$, at several fixed values of $M^2$.\\\\
{\bf Fig. (4)} The same as Fig. (3), but at the fixed value of the
continuum threshold $s_0=4.5\,GeV^2$.

\newpage

\begin{figure}
\vskip 3. cm
    \includegraphics{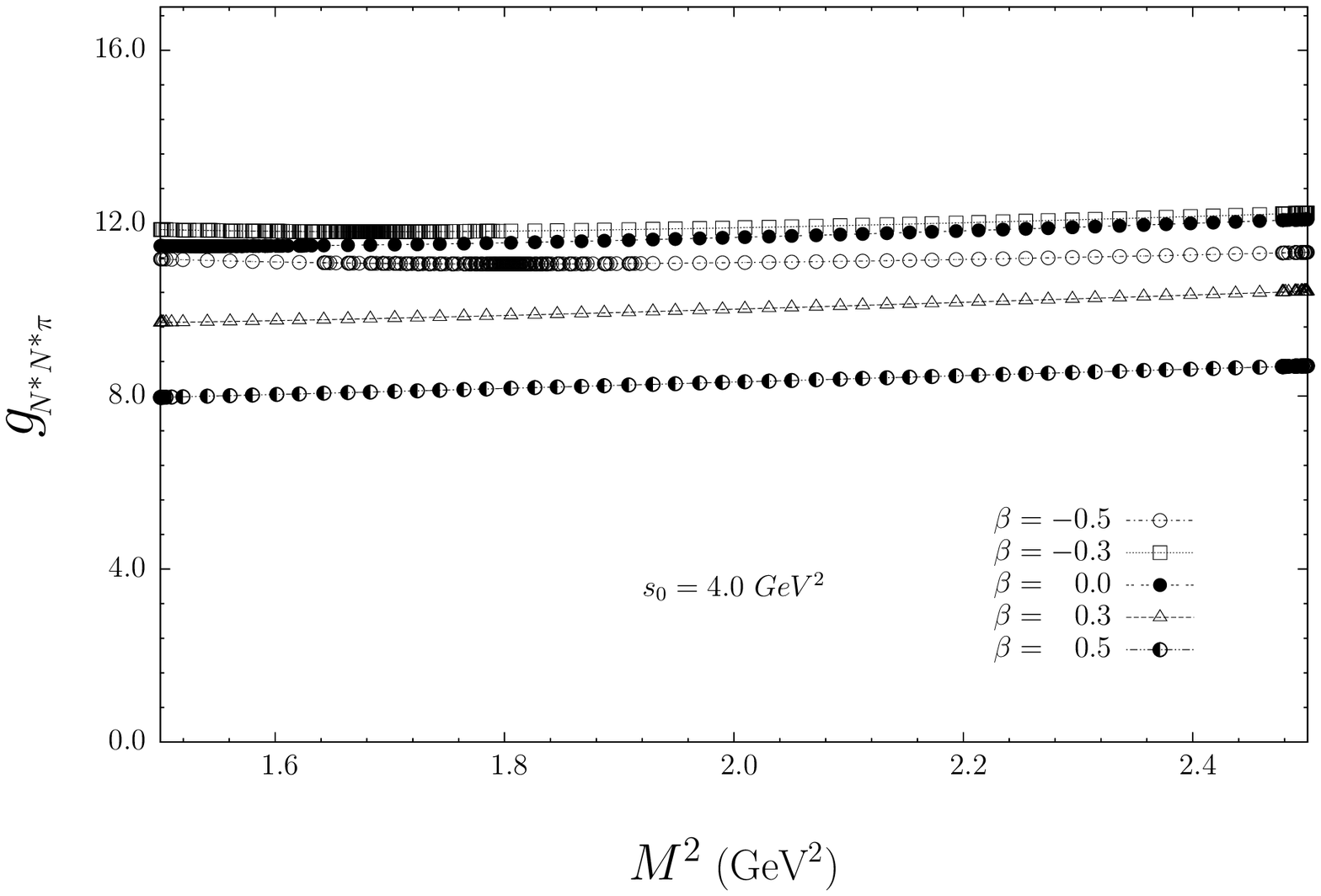}
\vskip 7.0cm
\caption{}
\end{figure}

\begin{figure}
\vskip 3. cm
    \includegraphics{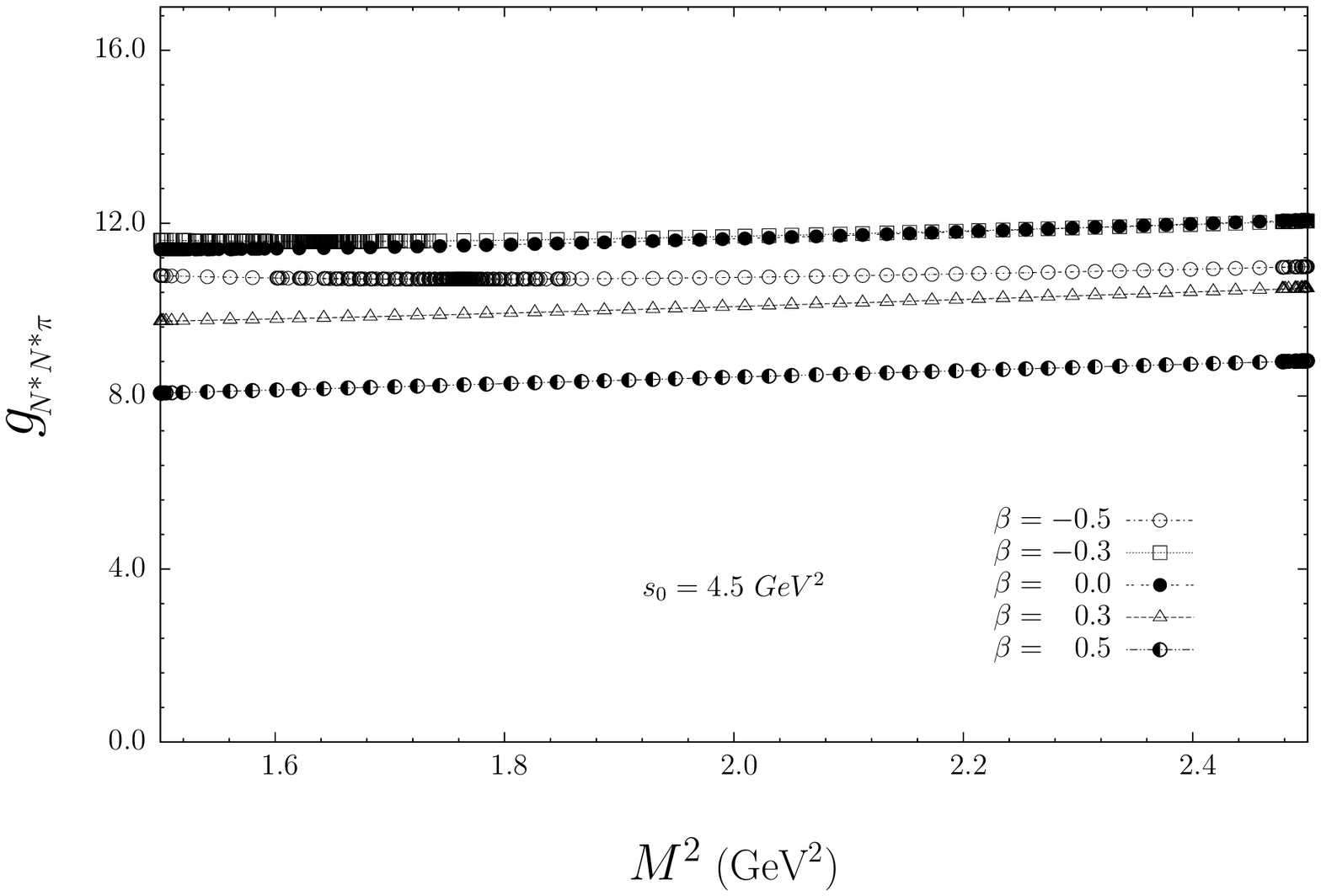}
\vskip 7.0cm
\caption{}
\end{figure}

\begin{figure}
\vskip 3. cm
    \includegraphics{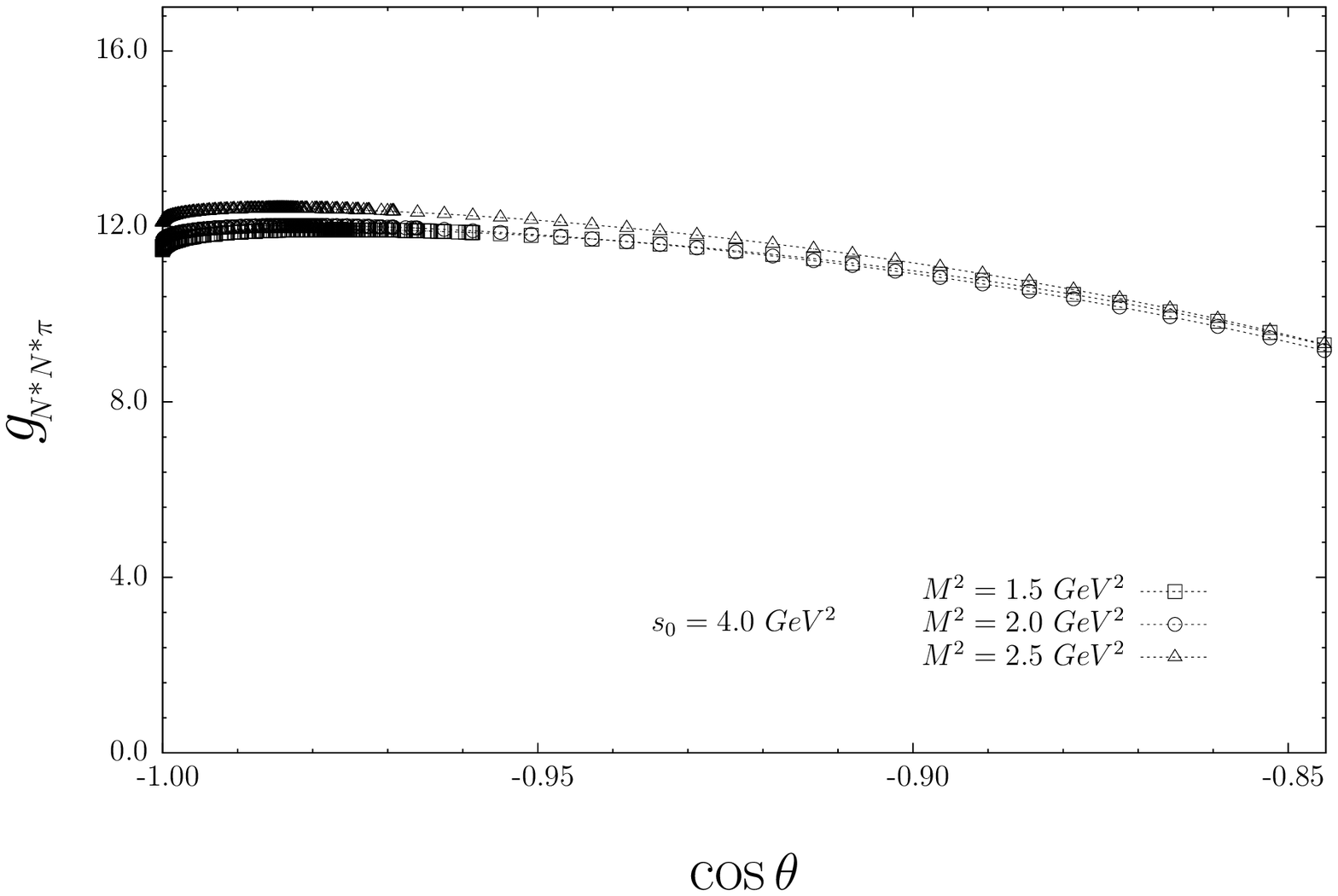}
\vskip 7.0cm
\caption{}
\end{figure}

\begin{figure}
\vskip 3. cm
    \includegraphics{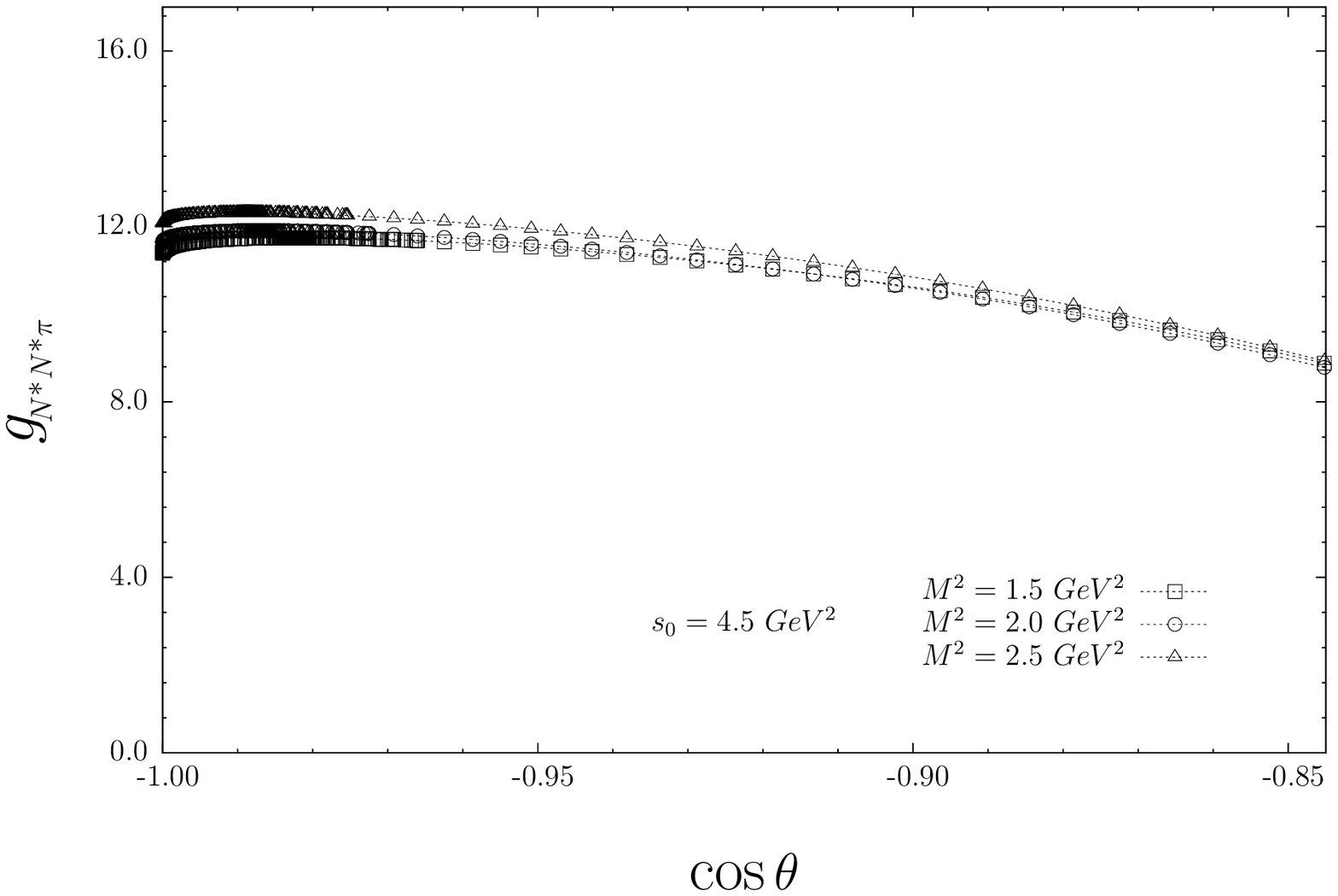}
\vskip 7.0cm
\caption{}
\end{figure}

\end{document}